\documentstyle[twoside,fleqn]{article}

\newcommand{\bls}[1]{\renewcommand{\baselinestretch}{#1}}

\def\noi{\noindent}

\makeatletter
\renewcommand{\section}{\@startsection{section}{1}{0pt}%
        {-3.5ex plus -1ex minus -.2ex}{2.3ex plus .2ex}%
        {\large\bf\protect\raggedright}}

\renewcommand{\subsection}{\@startsection{subsection}{2}{0pt}%
        {-3ex plus -1ex minus -.2ex}{1.4ex plus .2ex}%
        {\normalsize\bf\protect\raggedright}}

\renewcommand{\thesubsubsection}%
        {\arabic{section}.\arabic{subsection}.\arabic{subsubsection}.}
\renewcommand{\@oddhead}{\raisebox{0pt}[\headheight][0pt]{%
   \vbox{\hbox to\textwidth{\rightmark \hfil \rm \thepage \strut}\hrule}}}
\renewcommand{\@evenhead}{\raisebox{0pt}[\headheight][0pt]{%
   \vbox{\hbox to\textwidth{\thepage \hfil \leftmark \strut}\hrule}}}
\newcommand{\heads}[2]{\markboth{\protect\small\it #1}{\protect\small\it #2}}
\newcommand{\Acknow}[1]{\subsection*{Acknowledgement} #1}
\makeatother

\newcommand{\Title}[1]{\noindent {\Large #1} \\}
\newcommand{\Author}[2]{\noindent{\large\bf #1}\\[2ex]\noindent{\it #2}\\}
\newcommand{\Abstract}[1]{\vskip 2mm \begin{center}
     \parbox{16.4cm}{\small\noindent #1} \end{center}\bigskip}
\newcommand{\foom}[1]{\protect\footnotemark[#1]}
\newcommand{\foox}[2]{\footnotetext[#1]{#2}}
\newcommand{\email}[2]{\footnotetext[#1]{e-mail: #2}}

\newcommand{\sect}[1]{Sec.\,#1}

\def\nq{\hspace{-1em}}
\def\nqq{\hspace{-2em}}
\def\nhq{\hspace{-0.5em}}

\def\cm{\hspace{1cm}}
\def\inch{\hspace{1in}}

\newcommand{\Eq}[1]{Eq.\,(\ref{#1})}

\def\beq{\begin{equation}}
\def\eeq{\end{equation}}
\def\bear{\begin{eqnarray}}
\def\al{&\nhq}
\def\lal{&&\nqq {}}               
\def\bearr{\begin{eqnarray} \lal}
\def\ear{\end{eqnarray}}
\def\earn{\nonumber \end{eqnarray}}

\def\nn{\nonumber\\ {}}

\def\nnn{\nonumber\\ \lal }

\def\yy{\\[5pt]}
\def\yyy{\\[5pt] \lal}
\def\eql{\al =\al}

\def\e{{\,\rm e}}

\def\sign{{\,\rm sign\,}}

\def\dim{{\,\rm dim\,}}
\def\const{{\rm const}}

\def\then{\ \Rightarrow\ }

\newcommand{\vars}[1]{\left\{\begin{array}{ll}#1\end{array}\right.}
\newcommand{\lims}[1]{\mathop{#1}\limits}

\def\wider{\vphantom{\int}}

\def\umx{u_{\max}}
\def\uh{u_{\rm hor}}
\def\uinf{u_{\infty}}
\def\sumi{\sum_{i=1}^{n}}

\def\ph{\varphi}
\def\conf{\rm conf}
\def\xic{\xi^{\conf}}

\def\R {{\cal R}}
\def\A {{\cal A}}
\def\B {{\cal B}}
\def\go{\overline{g}{}}
\def\Ro{\overline{\R}{}}
\def\ds{ds_D^2}
\def\dso{d\overline{s}^2}
\def\Vo {\overline{V}{}}
\def\omo{\overline{\omega}}
\def\phim{$\ph^{\min}$}
\def\phic{$\ph^{\conf}$}
\def\Lm{L_{\rm m}}

\def\sph{spherically symmetric\ }
\def\bh{black hole}
\def\wh{wormhole}

\heads{K.A. Bronnikov}
{Extra Dimensions, Nonminimal Couplings, Horizons and Wormholes}

\bls{1.01}

\begin{document}
\thispagestyle{empty}
\twocolumn[
\noi \unitlength=1mm
\begin{picture}(174,8)
   \put(31,8){\shortstack[c]
       {RUSSIAN GRAVITATIONAL SOCIETY                \\
       INSTITUTE OF METROLOGICAL SERVICE             \\
       CENTER OF GRAVITATION AND FUNDAMENTAL METROLOGY}       }
\end{picture}
\begin{flushright}
                                         RGS-VNIIMS-96/06 \\
                                         gr-qc/9703020 \\
	{\it Grav. and Cosmol.} {\bf 2}, 3\foom 1, 221-226 (1996)
\end{flushright}
\bigskip

\Title{EXTRA DIMENSIONS, NONMINIMAL COUPLINGS,\yy
	  HORIZONS AND WORMHOLES}

\Author{K.A. Bronnikov\foom 2}
{Center for Gravitation and Fundamental Metrology, VNIIMS,
     3--1 M. Ulyanovoy Str., Moscow 117313, Russia}

{\it Received 20 April 1996} \\

\Abstract
	{Static, \sph configurations of gravity with nonminimally coupled
	scalar fields are considered in $D$-dimensional space-times ($D\geq 4$)
	in the framework of generalized scalar-tensor theories (STT).
	We seek special cases when the system has no naked
	singularity but, instead, forms either a \bh, or a \wh. General
	conditions when this is possible, are formulated. In particlar, some
	such special cases are indicated for multidimensional Brans-Dicke
	theory and for linear, massless, nonminimally coupled scalar fields
	(the coupling $\xi \R \ph^2$ where $\xi=\const$ and $\R$ is the
	curvature).  It is shown that in the Brans-Dicke theory the only \bh\
	solution corresponds to $D=4$ and the coupling constant $\omega< -2$,
	and there is a \wh\ solution corresponding to $\omega=0$. For the
	$\xi$-coupled linear scalar field it is shown that the only \bh\
	solution is the well-known one, with a conformal scalar field
	($\xi=\xic=1/6$)  in 4 dimensions (a \bh\ with scalar charge), while the
	known 4-dimensional \wh\ solution is generalized to systems
	with conformal coupling in arbitrary dimension.}

] 
\foox 1 {This number of the journal is dedicated to the 80th birthday of
         Prof. K.P. Staniukovich (1916--1989).}
\email 2 {kb@goga.mainet.msk.su}

\section{Introduction}                 

	Prof. K.P. Staniukovich believed that general relativity (GR) is not an
	ultimate theory of gravity even on the classical level and paid much
	attention to its generalizations \cite{stan,hydro}. This paper,
	submitted to an issue dedicated to his memory, touches upon some
	specific problems on this trend.

	As is well-known, in GR all \sph scalar-vacuum and
     scalar-electrova\-cu\-um configurations possess naked singularities
     if the scalar field is massless, minimally coupled (\phim). Their
     counterparts with a conformally coupled scalar field
	(\phic) provide a wider spectrum of possibilities: in the
	general case there are naked singularities as well, but more various
	types of these, and, moreover, in some special cases they
	describe \bh s \cite{boch,bek} or \wh s \cite{br-acta}.

	A natural question arises: are these new possibilities a distinctive
	feature of conformal coupling and/or the 4-dimensional nature of the
	space-time, or they occur for more general nonminimal scalar field
	couplings in various dimensions?

     GR with \phim\ and \phic\ are special cases of a rather
	general model called generalized scalar-tensor theory (STT) of
	gravity (see e.g. \cite{wag}), with the Lagrangian
\beq
	L = \A(\ph){\R} + \B(\ph)\ph^A\ph_A - 2\Lambda(\ph)+\Lm   \label{Lagr}
\eeq
	where $\R$ is the scalar curvature of a Riemannian space-time $V^D$ of
	arbitrary dimension $D$; $\A,\ \B$ and $\Lambda$ are
	functions varying from one specific STT to another and $\Lm$ is the
	nongravitational matter Lagrangian.  One frequently considers GR with
	linear scalar fields, which form a special case of (\ref{Lagr}), such
	that
\beq
	\A(\ph) = 1 - \xi \ph^2, \qquad \B \equiv 1,          \label{Nonmin}
\eeq
	where $\xi$ is the nonminimal coupling constant.
	In particular, $\xi =0$ corresponds to minimal coupling
	and $\xi=\xic = (D-2)\big/[4(D-1)]$ to conformal coupling.

	There are many reasons for considering nonminimal couplings.
	Historically, STT were put forward as viable theories other
	that GR, able to account for the observed effects of relativistic
	gravity or slightly modify the corresponding predictions of GR
	\cite{jordan,bdicke}. In modern theory of the early Universe, STT
	are one of the ways to create successful inflationary models
	(``extended inflation'', \cite{infla} and many others). In quantum
	field renormalization theory the constant $\xi$ appears as a free
	parameter to be determined empirically; a closely related issue is the
	induced gravity concept (\cite{zee} and others), where gravity itself
	essentially results from nonminimal scalar field coupling.  Other
	sources of nonminimal couplings are the modern unification theories ---
	(super)string and Kaluza-Klein ones.  On the other hand, properties of
	\sph configurations are one of the key issues in any theory of gravity.

	Returning to the above question on the occurrence of horizons and \wh s
	in models more general than 4-dimensional GR with \phic,
	for a very narrow range of generalizations an answer was given in
     Ref.\,\cite{Xan}:  it was shown that static scalar-vacuum solutions of
     $(d+2)$-dimensional GR, with \phic\ and the space-time
	structure $U^2 \times S^d$ (where $U^2$ accounts for the radial and
	temporal variables), can possess horizons only when $d=2$, i.e., in the
	conventional 4-dimensional case.

	Here we consider the broad class of theories (\ref{Lagr})
	in the case $\Lambda = \Lm =0$, for which
	a general exact static, \sph solution is available. The space-time
	structure and the metric are assumed in the form
\bearr
     V^D (g) =
	U^2 \times S^2 \times V_1 \times\cdots\times V_n, \nnn   \label{Stru}
     \qquad \dim M_i=N_i;     \qquad      D=4+\sumi N_i,
\ear
     where $V_i\ (i=1,\ldots,n)$ are Ricci-flat internal spaces
	of arbitrary dimensions $N_i$ and signatures;
\bearr                                                           \label{DsD}
     \ds = \e^{2\gamma}dt^2 - \e^{2\alpha}du^2 - \e^{2\beta}d\Omega^2
 		+\sumi \e^{2\beta_i(u)}ds_i^2;                         \nnn
\ear
	$d\Omega^2 = d\theta^2 + \sin^2\theta\ d\phi^2$ and $ds_i^2$ are the
	linear elements on $S^2$ and $V_i$, respectively.

	In \sect 2 we write out the relevant scalar-vacuum solution.
	In \sect 3 we try to find general conditions when this solution can
	contain horizons or describe a \wh. Further on the study is
	specialized to $D$-dimensional Brans-Dicke theory (\sect 4) and
	GR with a linear, nonminimally coupled scalar field (\sect 5).
	\sect 6 is a brief conclusion.

\section{Vacuum solutions of generalized scalar-tensor theories}    

     The system (\ref{Lagr})  is essentially reduced to that with \phim\
     by a conformal mapping well-known in 4-dimen\-sional STT \cite{wag}
	and modified for $D$ dimensions as follows \cite{birk}:
\beq                      \wider                                     
	V^D(g) \to \Vo^D (\go): \qquad
					g_{MN}= A^{-2/(D{-}2)}\go_{MN}.	\label{Conf}
\eeq
	Indeed, omitting a total divergence, we obtain the following form
	of the Lagrangian in terms of $\go$:
\bearr                                                                
	\overline{L} = \Ro + F(\ph)\go^{AB}\ph_A\ph_B           \nnn
    \inch 	+ A^{-D/(D{-}2)}[- 2\Lambda(\ph)+\Lm]       \label{Lagr1}
\ear
	where an overbar marks quantities corresponding to $\go_{AB}$ and
\beq                                                       \label{Fph}
	F(\ph) = \frac{1}{\A^2}\biggl[\A\B +                             
		\frac{D-1}{D-2}\biggl(\frac{d\A}{d\ph}\biggr)^2\biggr].
\eeq
	Putting in (\ref{Lagr}) $\Lambda = \Lm =0$,
	it is possible to write down the general static, \sph
	scalar-vacuum solution to the field equations in the following form
	\cite{br-iv,bm-itogi}:
\bearr
	\ds = f(u)\dso,   \cm       f(u) = [\A(\ph)]^{-2/(D-2)};   \nnn
\	\dso = \e^{-2b_0 u}dt^2
	          -\frac{e^{2Bu}}{s^2(k,u)}
		  	  \biggl[\frac{du^2}{s^2(k,u)}+ d\Omega^2\biggr] \nnn
\cm\cm\cm     + \sumi \e^{-2b_i u}ds_i^2,                      \nnn  
	F(\ph)(d\ph/du)^2 = S = \const                  \label{Sol-u}
\ear
     where $u$ is the radial coordinate, harmonic in $\Vo$
	(such that  $\overline{\nabla}^M \overline{\nabla}_M u =0 $),
     defined for $u>0$; the flat-space asymptotic corresponds to $u=0$;
	the integration constants $B$, $b_i$, $k$ and $S$ are connected by the
	relations
\bearr
	B = b_0 + \sumi N_i b_i; \nnn                                    
	2k^2\sign k = B^2 + b_0^2 + \sumi N_i b_i^2 +S;       \label{Int-u}
\ear
	lastly, the function $s(k,u)$ is defined as follows:
\beq                                                                 
	s(k,u) \equiv \vars{ (1/k)\sinh ku,  \quad & k>0,\\
			      			   u, &        k=0,\\
				      (1/k) \sin ku,  &        k<0.}    \label{Def-s}
\eeq
	The constant $S$ has the meaning of a scalar charge; with $S=0$ we are
	led to $D$-dimensional GR. We will be naturally interested in the
	nontrivial case $S\ne 0$.

	As follows from the last line of (\ref{Sol-u}), due to $S=\const$
	the function $F(\ph)$ has the same sign in the whole space
	(or at least in the $u$-chart which includes the asymptotic region).
	Therefore,
	applied to our \sph case, all STT are divided into two large classes:
	$F(\ph)>0$ and $F(\ph) <0$, hereafter labelled as {\bf normal} and
	{\bf anomalous}, respectively. In normal STT the gradient term in
	(\ref{Lagr1}) has its conventional sign and consequently the scalar
	field energy density is nonnegative in this conformal frame.

	By (\ref{Int-u}) we have $k>0$ for all normal STT.
	Thus many possible solution behaviours, connected with
	$k<0$, are possible only in anomalous STT with $F<0$.

	An alternative form of $\dso$ for $k>0$ is obtained after the
	coordinate transformation
\beq                                                                 
	\e^{-2ku} = 1-2k/R \equiv P(R),                     \label{u-to-R}
\eeq
	namely,
\bearr                                                              
 \nq	\dso = P^{a_0}dt^2 - P^{-A}dR^2 - P^{1-A}R^2 d\Omega^2
			+ \sumi P^{a_i} ds_i^2       \nnn         \label{Sol-R}
\ear
	where the constants $k$, $a_i = b_i/k$, $A=B/k$ and $S$ are connected
	by the relation
\beq                                                                
	A^2 + a_0^2 + \sumi N_i a_i^2 + S/k^2 = 2.          \label{Int-R}
\eeq

\section{Search for horizons and wormholes}             

\subsection{Criteria}                                   

	We will try to find special cases when the metric
	$ds_D^2$ describes a \bh, i.e., possesses an event horizon. That means,
	in terms of (\ref{DsD}), that at some value of the radial coordinate
	($u=\uh$)
\begin{description}
\item[A1.]      $\e^{\gamma}\to 0$, while
\item[A2.]      $\e^{\beta}$ remains finite and
\item[A3.]      $\e^{\beta_i}$ $(i=\overline{1,n})$ remain finite;
\item[A4.]      No signal can reach $u=\uh$ for a finite time by a
	remote observer's clock, i.e., the integral $\int du\e^{\alpha-\gamma}$
	diverges as $u \to \uh$.
\end{description}
	In the formulation of Items A1-A4 the radial coordinate is arbitrary,
	not necessarily harmonic.

	We will also look for cases when the solution describes a (static,
	traversable) \wh, i.e., there are two flat-space asymptotics
	connected by a regular bridge. That
	means that, as well as at $u=0$, at some other value of the radial
	coordinate $u=\uinf$
\begin{description}
\item[B1.]  $\e^{\gamma}$ and
            $\e^{\beta_i}$ \   $(i=\overline{1,n})$ remain finite;
\item[B2.]  $\e^\beta \to \infty$;
\item[B3.]  There is an infinite path along the radius, i.e., the integral
            $\int \e^\alpha du$ diverges;
\item[B4.]  A correct flat-space circumference-radius ratio for
            coordinate circles is asymptotically valid, i.e.,
		  $\e^{\beta-\alpha}\beta' \to 1$.
\end{description}
	These criteria are also radial coordinate reparametrization invariant.

\subsection{Search for horizons ($k > 0$)}                 

	The solution (\ref{Sol-u}) is regular for $u < \infty$ ($R > 2k$),
	provided the function $f(u)$ is regular (vanishing or blowing-up of
	$f(u)$ at finite $u$ can lead only to a naked singularity). So a
	horizon can exist either at the sphere $u=\infty$ ($R=2k$), or somewhere
	beyond this sphere if the latter is regular. Consider the
	first possibility in terms of (\ref{Sol-R}).

	Criterion A3 implies that all $a_i$ ($i \geq 1$) are equal, and with no
	loss of generality\footnote{In what follows, for the same reason
	in all relevant cases we adopt the same assumption.}
	we will assume that there is only one internal space
	$V_1$, with $ \dim V_1 = N_1 = N$; as also follows from A3,
\beq                                                                
	f(u) \sim P^{-a_1} \quad {\rm as } \quad R\to 2k.      \label{I}
\eeq
	From A1 and A2 it follows
\bearr
     a_0 > a_1,                                   \label{II}   \yyy 
	a_0 + (N+1) a_1 =1,                          \label{III}       
\ear
	respectively. Finally, A4 leads to  $A+a_0 \geq 2$, whence
\beq
	a_1 (N + 2) \leq 0.                         \label{IV}
\eeq
	On the other hand, \Eq{Int-R} with (\ref{III}) gives
\beq
	S/k^2 = a_1 (N+2)[2- (N+1) a_1].              \label{V}
\eeq
	By (\ref{IV}), $a_1 \leq 0$; but $a_1 =0$ with (\ref{V}) leads to
	$S=0$, i.e., the trivial case $a_0=A=1$, $S=0$, that is, $\ph=\const$
	and the Schwarzschild metric with ``frozen'' extra dimensions (the only
	\bh\ solution in the minimal coupling case $F(\ph)=\const$
	\cite{br-iv}).

	The other option, $a_1 <0$, leads, by (\ref{V}), to $S<0$ and
	consequently to $F <0$. We arrive at

\proclaim  Proposition 1. Nontrivial (non-Schwarzschild) \bh s with
	horizons at $u=\infty$ can exist only in anomalous STT.

	Moreover, in this case
\bearr
	f(u) \equiv [\A(\ph)]^{-2/(D-2)} \sim P^{-a_1} \to 0 \nnn
	\quad \A(\ph) \to \infty  \quad {\rm as}\quad u \to \infty, \label{VI}
\ear
	i.e., the effective gravitational coupling ($\sim 1/\A$) vanishes at
	the horizon.

	If $u=\infty$ is a regular sphere, the solution behaviour
	depends on $\A(\ph)$ and cannot be determined in a general
	manner; however, the known example of \bh s with a conformal scalar
	field in 4 dimensions makes sure that horizons beyond such a sphere
	are, in principle, possible.

	This sphere is regular if all the metric coefficients in (\ref{DsD})
	are finite, i.e., all $a_i$ are equal and, in addition,
\beq
	a_i = 1 -A \  \then  \ a_i = 1/(N +2).                \label{VII}
\eeq
	Then \Eq{Int-R} implies
\beq
	S/k^2 = 1 + 1/(N+2) = 1/(4\xic).                      \label{VIII}
\eeq
	The result $S>O$ means that $F(\ph) > 0$. So we have proved

\proclaim Proposition 2.
	A continuation beyond $u=\infty$ is possible only in normal STT.

	Moreover, $f(u) \sim P^{-a_0} \to \infty$ as $ u\to \infty$, i.e., the
	effective gravitational coupling tends to infinity. The experience of
	dealing with such a behaviour in 4 dimensions \cite{br-kir}
	indicates that a strong gravitational instability can probably develop
	near such a sphere.

	We also notice that in the present case the functions $g_{aa}(u)$
	($a>0$) and $g_{00}(u)$ coincide up to a constant scale factor. This
	coincidence will be naturally preserved beyond $u = \infty$ as well.
	Therefore Criteria A1 and A3 cannot be fulfilled together. We
	can conclude the following:

\proclaim Proposition 3.	Event horizons beyond the sphere $u=\infty$ are
	possible only in the case $D=4$.

\subsection{Search for horizons ($k \leq 0$)}                 

	In the case $k=0$ the solution is regular at $u < \infty$ provided
	$f(u)$ is regular. As $u\to \infty$, $\e^{\beta}$ behaves like
	$\e^{\const\cdot u}/u$, while $\e^{\beta_1}$ behaves like just
	$\e^{\const\cdot u}$. Therefore $g_{22}$ and $g_{aa}$ ($a>3$) cannot
	simultaneously tend to constants and for $D>4$ the surface $u=\infty$
	can be neither a regular sphere, nor a horizon.

	If $k=0,\ D=4$, a horizon at $u=\infty$ is possible if
\beq
	b_0 > 0, \qquad f(u) \sim u^2 \e^{-2b_0 u}.           \label{IX}
\eeq
     Then $g_{00} \sim u^2\e^{-4b_0 u} \to 0$ and A4 is also valid. So
	{\bf this is a possible \bh\ case}.

	Let us now address to the case $k<0$, so that in (\ref{Sol-u})
	$s(k,u) = k^{-1}\sin ku$ and the solution is defined for
	$0 < u < \umx = \pi/|k|$. All the exponential functions are finite. So,
	in the general case, when $f(u)$ is regular for $0 \leq u \leq \umx$,
	the solution,	as is easily verified,
	describes {\bf a wormhole} \cite{br-acta,hellis}.
	If, however, $f(u)\to 0$ as $u\to\umx$, so that $g_{22}$ be finite,
	then at the same time $g_{00} \to 0$ and $g_{aa}\to 0$ ($a>3$), in
     contrast to Criterion A3. Thus, in addition to Proposition 3, we have

\proclaim Proposition 4. Event horizons for $k<0$ can
     exist only if $D=4$.

     Criterion A4 is then valid as well and, in
	addition, $\A \to \infty$ as $u\to\umx$, so that again the effective
	gravitational coupling blows up.

\subsection{Search for \wh s}                                

	We have seen that for $k<0$ (that is, only in anomalous STT) \wh s
	appear in the general case. Let us find out when they are possible in
	normal STT.

	They are evidently possible when the space-time is continued beyond
	$u=\infty$, as shown by an explicit example of this sort \cite{br-acta}
	($D=4$, GR with \phic).

	Another possibility is that the second flat asymptotic occurs at
	$u=\infty$ ($R=2k$). In this case one must have in (\ref{Sol-R})
\beq
	a_0=a_1, \qquad  f(u)
     \mathop{\sim}\limits_{R \to 2k} P^{-a_0}.   \label{X}
\eeq
	\Eq{Int-R} then takes the form
\beq
	(N+1)(N+2) a_0^2 + S/k^2 =2.                            \label{XI}
\eeq
	On the other hand, Criteria B2 and B3 lead to
\beq
	a_0 \geq 2/(N+2),                                       \label{XII}
\eeq
	whence by (\ref{XI})
\beq
	S/k^2 \leq - 2N,  \then    F <0,                        \label{XIII}
\eeq
	i.e., this situation is possible only for anomalous STT. Moreover, the
	asymptotic circumference-to-radius ratio is $\sim P^{-1/2}\to \infty$,
	i.e., Criterion B4 is violated: this configuration is not a \wh.
	A conclusion is:

\proclaim Proposition 5. In normal STT \wh s are possible only with a
	continuation beyond $u=\infty$.

\section{Example: the Brans-Dicke theory}                 

	A multidimensional generalization of the Brans-Dicke STT is specified
	by the functions
\beq                                                               
	\A(\ph) = \ph,  \qquad \B(\ph)=\omega/\ph,
		\qquad \omega  = \const.                           \label{XIV}
\eeq
	Consequently,
\bear                                                              
	F(\ph) \eql \omo / \ph^2, \qquad \omo = \omega+ \frac{D-1}{D-2}, \nn
	f(u)    \eql \ph^{-2/(D-2)},								  \nn
	F\ph^2 \eql S        \then    \ph=\ph_0 \e^{su}      \label{XV}
\ear
	where $s=\const = \pm \sqrt{S/\omo}$. In what follows we omit the
	unessential constant $\ph_0$. Recall that the metric is given in
	(\ref{Sol-u}) or, for $k>0$, (\ref{Sol-R}).

	Let us first assume that the STT is normal, i.e., $\omo >0$. As follows
	from the above considerations, \bh s or \wh s are then possible only
	beyond a regular sphere $u=\infty$ ($R=2k$). One easily finds that this
	sphere can be regular if
\bearr                                                             
	a_0 = a_1 = 1/(D{-}2); \qquad S/k^2 = (D{-}1)/(D{-}2) \nnn \label{XVI}
\ear
	and the function $f(u)$ takes the form
\bearr                                                               
	f(u) = \e^{-2su/(D{-}2)} \equiv P^{c},
		\qquad c=\pm \sqrt{\frac{D{-}1}{(D{-}2)^3\, \omo}}.
										\nnn \label{XVII}
\ear
	Then the regularity condition $f(u)P^{a_0}\to \const$ as $u\to\infty$
	implies that in (\ref{XVII}) the minus sign must be chosen and
\beq                                                               
	\omo = (D{-}1)/(D{-}2)\ \ \then \ \ \omega =0.           \label{XVIII}
\eeq
	Thus the continuation is possible only in the special case of the STT
	(\ref{XIV}) with $\omega=0$. Under the above conditions the metric
	takes the simple form
\beq                                                               
	\ds = dt^2 - \frac{dR^2}{1-2k/R} -R^2 d\Omega^2 +ds_1^2.
\eeq
	The continuation beyond $R=2k$ is realized, for example, by putting
	$x= \sqrt{R-2k}$ and allowing $x$ to take all real values. Then
\bearr                                                             
	\ds = dt^2 -4(x^2+2k)dx^2 - (x^2+2k)^2 d\Omega^2 + ds_1^2,    \nnn
\ear
	i.e., a wormhole, manifestly symmetric under the substitution
	$x\to -x$, with the neck radius $\sqrt{2k}$, trivial extra dimensions
	and zero mass (since $g_{00}=\const$). However, the scalar field is
	nontrivial: $\ph = x/\sqrt{x^2{+}2k}$.

	In the anomalous case $\omo <0$ there is no continuation beyond
	$u=\infty$ and the only nontrivial possibility is that of a horizon at
	$u=\infty$ for $k>0$. Indeed, for $k<0$, when $u$ is defined on a
	finite segment, the exponential conformal factor $f(u)$ cannot change
	the metric qualitatively; for $k=0$, $g_{22}$ behaves like
	$e^{\const \cdot u}/u$ and cannot tend to a finite limit as $u\to
	\infty$.

	For $k>0$ the requirements A1--A4 with $a_1\ne 0$ are fulfilled if and
	only if $\omega < -2$ and
\bearr                                                              
	a_0 = 1-(N+1)a_1; \qquad a_1 = \frac{2}{(N+2)(\omega+2)} <0, \nnn
\ear
	so that the only remaining free integration constant is $k$,
	connected with the \bh\ mass. The ``scalar charge'' $S$ is
	also expressed in terms of $k$ and $\omega$:
\beq
	S= \frac{4}{\omega +2}\biggl[1 + \frac{N+1}{(N+2)|\omega +2|}\biggr].
\eeq

	The assumption $a_1=0$ leads to a Schwarzschild \bh\ with trivial extra
	dimensions.

	However, the case $D=4$, when the condition A3 is
	cancelled, must be considered separately. It turns out that a
	non-Schwarzschild horizon at $u=\infty$ indeed takes place
	if and only if $\omega <-2$ and
\beq
	a_0 = \frac{|\omega|-1}{|\omega|-2} >1,\qquad S/k^2 = 2(1-a_0^2).
\eeq
	There is again only one free integration constant $k$.

\section{Linear, nonminimally coupled scalar fields} 

\subsection{The general case}                        

	We have seen that there are indeed some cases
	when the conformal factor $f(u)$ induces new qualitative
	features of the solution behaviour, the existence of \bh\ and \wh\
	configurations. Let us now consider probably the most interesting
	special case of STT, namely, that described by \Eq{Nonmin}. In this
	case, in \Eq{Fph})
\beq
	F(\ph) = \frac{1-\eta \ph^2}{(1-\xi \ph^2)^2}, \qquad
				   \eta = \xi(1-\xi/\xic).
\eeq

	We assume $\xi\ph^2 < 1$, i.e., the values of $\ph$ when the STT
	is normal at the asymptotic (in the opposite case, the effective
	gravitational coupling would be negative, which seems nonphysical).
	Then the possibility of \bh s or \wh s may be connected only with a
	continuation beyond $u=\infty$.

	The general consideration of Sec.\,3 implies that, for
	naked-singularity-free solutions, a regular sphere $u=\infty$ and
	a continuation beyond it must be provided. For such a sphere
	$A \lims{\to}_{u\to\infty} 0$, therefore let us consider
	the asymptotic of $f(u)$ as $A\to 0$.

	Evidently, in this case $\xi > 0$, $\ph^2 \to 1/\xi$ and
	$1- \eta\ph^2 \to \xi/\xic$. Hence the last line of (\ref{Sol-u})
	gives:
\beq
	\sqrt{S} du \sim \frac{\xi}{\xic} \frac{d\ph}{1-\xi\ph^2}
\eeq
	which yields after integration
\beq                                                      \label{Xiphi}
	\sqrt{\xi}\ph = \tanh h(u+u_0), \qquad   h = \sqrt{S/\xi}\ \xic
\eeq
     where $u_0$ is an integration constant and ``$\sim$'' means the
	greatest term in a possible series decomposition kept in mind.
     From (\ref{Xiphi}) it follows
\beq
	f(u) \sim \e^{4hu/(D-2)}  \qquad (u\to \infty).
\eeq
	On the other hand, a regular sphere at $u=\infty$ can exist only
	if (see (\ref{u-to-R}) and (\ref{VII}))
\beq
	f(u) \sim \e^{2ku/(D-2)}.
\eeq
	Comparing the two expressions for $f(u)$ and excluding $h$, one obtains
\beq
	S/k^2 = \xi/(2 \xic)^2.                                \label{XIX}
\eeq
	Meanwhile, as follows from (\ref{VIII}), $S/k^2 = 1/(4\xic)$, which is
	compatible with (\ref{XIX}) only when $\xi=\xic$.

	We conclude that the continuation is possible only with a conformally
	coupled field, \phic. Therefore, although it is straightforward, we
	will not determine $f(u)$ explicitly in the general case.

\subsection{Conformal coupling}

	Let us now give an explicit form of the solution for $\xi=\xic$.
	One easily finds:
\bear                                                               
	\sqrt{\xi}\ph \eql \tan \bigl[\sqrt{\xi S}(u+u_0)\bigr]; \nn \label{XX}
	f(u) \eql \Bigl\{
		\cosh \bigl[\sqrt{\xi S}(u+u_0)\bigr] \Bigr\}^{4/(D-2)}.
\ear
	Under the above regularity conditions for the sphere $u=\infty$,
	$\sqrt{\xi S}=k/2$ and the corresponding special solution can be
	transformed from (\ref{Sol-u}) into
\bear                                                               
	\ds \eql \biggl[\frac{c}{2}
			\biggl(1+\frac{x}{c^2}\biggr)\biggr]^{4/(D-2)}  \nnn
    \nqq	\times	\biggl\{dt^2
	-\frac{4k^2}{(1-x^2)^2}\biggl[\frac{4dx^2}{(1-x^2)^2}+d\Omega^2\biggr]
                                            +ds_1^2 \biggr\},  \nn
	\ph (x) \eql \frac{1}{\sqrt{\xi}}\frac{c^2-x}{c^2+x},  \label{Sol-x}
\ear
	where we have introduced
\beq
	x = \e^{-ku}, \qquad  c= \e^{ku_0/2}.
\eeq
	So $x=1$ corresponds to spatial infinity, $x=0$ to the regular sphere
	$u=\infty$ and $x<0$ is the new domain uncovered by the $u$-chart.
	The solution behaviour turns out to strongly depend on the value of
	$c$, the constant determining the asymptotic value of the $\ph$ field:

\begin{description}
\item[1.]  $c<1$. \ At $x=-c^2 > -1$ the metric in (\ref{Sol-x}) has a
	naked singularity due to conformal factor vanishing.
\item[2.]
	$c=1$. \ The coordinate $x$ is defined up to $x=-1$. For $D=4$, after
	the further reparametrization
\beq
	k/(1-x)=r,
\eeq
	we obtain the old solution describing a \bh\ with a scalar charge
	\cite{boch,bek}
\end{description}
\bear                                                              
     ds^2 \eql \Bigl(1-\frac{k}{2r}\Bigr)^2 dt^2 -
              \Bigl(1-\frac{k}{2r}\Bigr)^{-2} dr^2 -r^2 d\Omega^2,  \nn
	\ph (r)\eql \frac{k}{\sqrt{\xi}(2r-k)}.                \label{Sol-r}
\ear
\begin{description}
\item[\phantom{xxxi}]
	For $D>4$, as $x\to -1$, the spherical radius $r=\sqrt{-g_{22}}$ tends
	to infinity. This is a kind of horizon, displaced infinitely far
	beyond a neck (a minimum of $r$), since both the proper length along
	the radial direction and the light travel time diverge as $x\to -1$.
\item[3.]
	$c>1$. The conformal factor before the curly bracket in (\ref{Sol-x})
	does not qualitatively affect the metric behaviour. One can easily
     verify that in this case $x\to -1$ is another flat-space asymptotic
     and all the \wh\ criteria B1--B4 are fulfilled. This \wh\ solution of
     arbitrary dimension generalizes that found in \cite{br-acta} for $D=4$.
\end{description}

\section{Conclusion}

     We have studied static, \sph scalar-vacuum configurations with both
     minimal and nonminimal scalar-metric couplings in space-times of
     arbitrary dimension. It has turned out that the introduction of
     nonminimal couplings can indeed widen the spectrum of solution
     behaviours, but for ``normal" STT (roughly, when the scalar field
     energy is positive-definite) something different from naked
     singularities, namely, \bh s or \wh s, takes place in very rare
     special cases. For instance, \bh s can appear only in 4 dimensions.
     For anomalous STT, apart from the general \wh\ case ($k<0$ in the
     solution), some special \bh\ configurations may exist.

     The situation was discussed in detail for the Brans-Dicke
     theory and the coupling $\xi R \ph^2$. For this $\xi$-coupling it
     has been shown that \bh s and \wh s appear only in the conformal
     case, $\xi=\xic$ (see (\ref{Nonmin})); the only \bh\
     solution is the well-known one ($D=4$, a \bh\ with scalar charge
     \cite{boch,br-acta,bek}, while the known 4-dimensional \wh\ solution
     \cite{br-acta} is generalized to arbitrary $D$.

	\Acknow
	{This work was supported in part by the Russian Ministry of Science.}

\end{document}